\documentstyle[11pt, paspconf, epsf]{article}
\begin{document}

\title{Warps and Secondary Infall}

\author{I. Garc\'{\i}a-Ruiz, K. Kuijken}
\affil{Kapteyn Astronomical Institute, Groningen}

\author{J. Dubinski}
\affil{Canadian Institute for Theoretical Astrophysics, Toronto}

\begin{abstract}
Secondary infall in galaxies could cause the angular momentum of
the outer halo to change its orientation. The generation of warps due to
this effect is studied using N-body simulations.
\end{abstract}

\section{Introduction}

In many spiral galaxies the outer regions of the disk curve away from
the symmetry plane of the inner disk, resembling an integral
sign. This is called a {\em warp}. The cause of this phenomenon is
still uncertain, but a dark halo may play an important role. Sparke \&
Casertano (1988) found some long-lived warping normal modes inside an
oblate halo potential, but they used a halo that was
unperturbed by the disk. Later, Dubinski \& Kuijken
(1995) and Nelson \& Tremaine (1995), following a comment by Toomre
(1983), showed that when the response of the halo is
taken into account the Sparke \& Casertano modes are no longer
long-lived.

Cosmic infall of matter onto galaxies has been proposed as a possible
explanation for the warp phenomenon (Ostriker \& Binney 1989). 
This infall likely causes the net angular momentum vectors of disk
galaxies to reorient each Hubble time.  

In this work, we simulate and analyze the response of a disk embedded
in a halo whose orientation is changed slowly. The properties of the
warps generated this way are studied later, both in shape and amplitude.

\section{Simulation details}

\subsection{Construction of the Halo - Disk models}

The halo and the disk have been constructed following Kuijken \&
Dubinski (1995), using a self-consistent approach. The halo has a
lowered Evans distribution function, and the disk is modeled with
concentric spinning rings. The mass profile and rotation curves of the
different models can be seen in figure [1]. The main parameters of
both models are summarized in table [1].

The number of particles used for each model was of
$100\hspace{0.75mm}000$ particles for the halo, and 400 rings, each of
them consisting of 90 particles.

\begin{figure}
\plotone{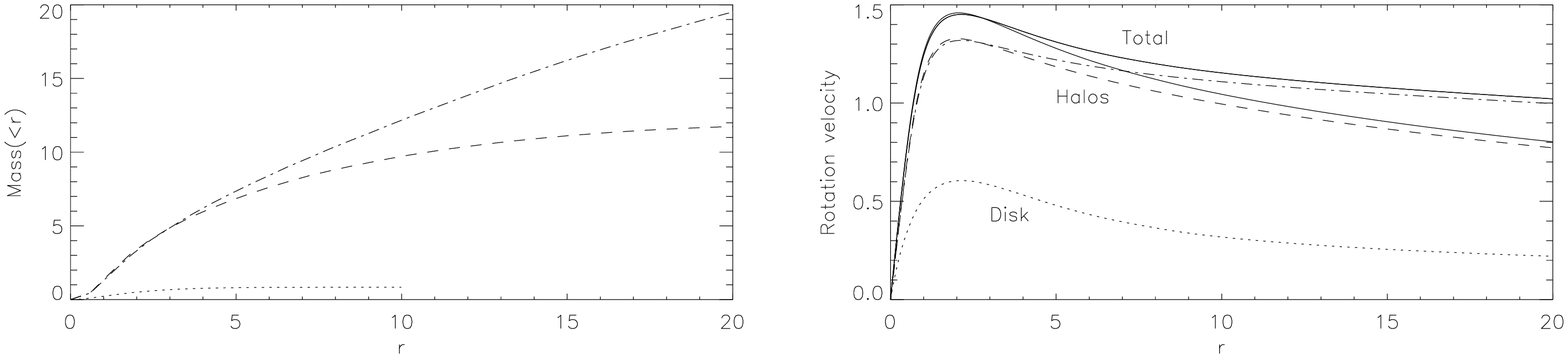}
\caption{{\small Mass profile and Rotation Curves: Disk (dotted), D
    model Halo (dash-dotted), and C model Halo (dashed)}}
\end{figure}

\begin{table}[h]
\begin{center}
\begin{tabular}{c@{\hspace{1cm}} c c c c c@{\hspace{1cm}}c c c c c}
  & \multicolumn{5}{c}{\small Disk} & \multicolumn{5}{c}{\small Halo} \\
  \small Model & \small $M_d$ & \small $R_d$ &  \small $R_t$ & \small $z_d$ & \small $R_{out}$ & \small ${\Psi}_0 $& \small ${\sigma}_0$ & \small $q$ & \small $C$ & \small $R_a $ \\ 
  \small       & \small (1)   & \small (2)   &  \small (3)   & \small (4)   & \small    (5)    & \small  (6)       & \small   (7)        & \small (8) & \small (9) & \small (10)   \\ \hline 
  \small  C    & \small 0.867 & \small  1    &  \small 8     & \small 0.1   & \small 0.5       & \small -6.0       & \small  1.32        & \small 0.9 & \small 0.1 & \small  0.8   \\
  \small  D    & \small 0.867 & \small  1    &  \small 8     & \small 0.1   & \small 0.5       & \small -7.0       & \small  1.30        & \small 0.9 & \small 0.1 & \small  0.8   \\ \hline 
\end{tabular}  
\end{center}
\caption{Model parameters: {\small (1) disk mass, (2) disk scale length, (3) disk truncation
  radius, (4) disk scale height, (5) disk truncation length, (6) halo
  central potential, (7) halo velocity dispersion, (8) halo potential
  flattening, (9) halo concentration, {$C={R_c}^2/{R_K}^2$}, (10)
  characteristic halo radius.}}
\end{table}

%These halo parameters give a Halo Mass of 12 and 28 (C and D models)

\subsection{Code used to evolve the system}

The N-body code use to evolve the system uses a hybrid approach: the
halo is made of particles, and its effect is computed with a Self-Consistent
Field code 
%(Hernquist, Sigurdsson \& Bryan, 1995)
(Hernquist \& Ostriker, 1992), expanding the
potential in terms of some basic functions. The disk, however, is
modeled using pivoted spinning rings. The dynamics on the disk are computed
calculating the potential and acceleration on the ``ring-particles'',
and then these are used to calculate the torque on each ring. This is
accomplished with a tree code (Barnes \& Hut, 1986).

\section{Results}

\begin{figure}
\plotone{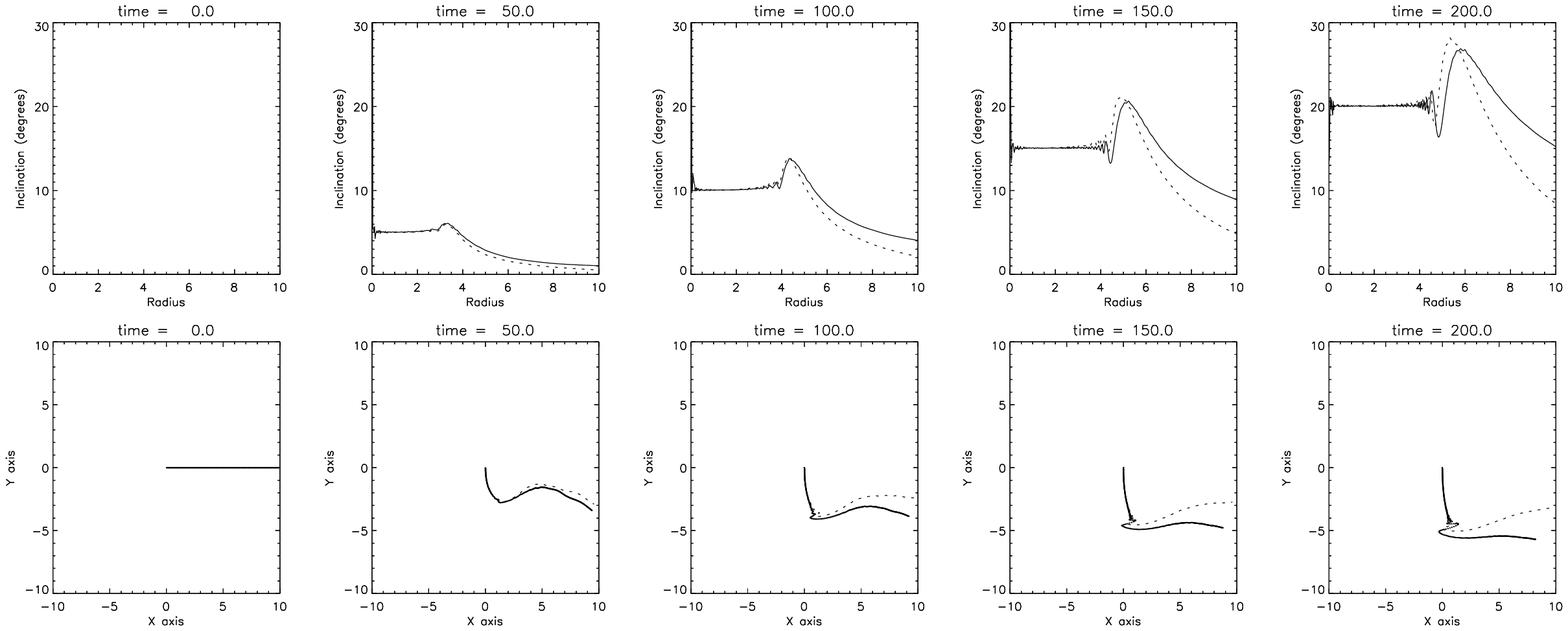}
\caption{Inclination angle (top) and orientation of the line of nodes
  (bottom) for
  models {\bf D} (solid) and {\bf C} (dotted). The inclination of the
  halo in this units is $\alpha=0.1time$ degrees}
\end{figure}

The first approach to the problem was to consider a rigidly rotating
halo, unaffected by the gravitational effect of the disk. The slewing
rate of the halo was chosen to be H$_0$ (100 km/s/Mpc for a disk scale
length of 3.5 kpc), so that the angular momentum
of the halo re-orients in the predicted timescale.The axis of
rotation of the halo lies in the initial halo-symmetry plane.

The results of the simulations for models {\bf C} and {\bf D} are
shown in figure [2]. Here the inclination of the rings and the position
angle of the line of nodes is plotted for different timesteps.

These plots show that the disk reorientation process
is fast enough to follow the halo in the inner parts. The disk warps
where its self-gravitation is not enough to hold it in the same
plane. This is in perfect agreement with the Briggs law (Briggs,
1990): straight line of nodes in the center, leading spiral
further out.
%The inner parts of the disk are the same for both models due to the
%fact that their halos have the same mass profile in the inner regions.
%Outside, model {\bf D} has more mass, so pulls more at the outer
%rings than in the model {\bf C} .
If the mass of the disk is doubled, the region aligned
with the symmetry plane of the halo increases (approximately one
scale length), as expected.  

\section{Next}

As mentioned in the introduction, the assumption of a non-responding
halo is very simplistic. The next step is to simulate the
re-orientation of a responsive halo that cosmic infall would cause. 
To do this we will put a rotating {\em tidal field} through 
the galaxy. This will make the symmetry plane of the outer halo change,
gaining angular momentum

This will be done by placing 2 {\em satellites} symmetrically, so that the
monopole term of the tidal field is zero, avoiding relative movements
of the galaxy with respect to the {\em satellites}. This will also
provide information about the influence of satellites on disks.

\end{document}